\documentclass[runningheads]{llncs}

\usepackage[T1]{fontenc}
\usepackage{graphicx}
\usepackage{multirow}
\usepackage{balance}
\usepackage{stfloats}
\usepackage{xcolor}
\usepackage{float}
\usepackage{url}
\usepackage{longtable}

\usepackage[table]{xcolor}
\usepackage{tikz}
\usepackage{diagbox}
\usetikzlibrary{positioning,fit,shapes}

\usepackage{tablefootnote}
\usepackage{listings}
\usepackage[edges]{forest}
\usepackage{tcolorbox}
\usepackage{mdframed}
\usepackage{fontawesome5}

\newmdenv[
  backgroundcolor=blue!9,
  leftline=true,
  rightline=false,
  topline=false,
  bottomline=false,
  linecolor=black,
  linewidth=3pt,
  innerleftmargin=10pt,
  innerrightmargin=10pt,
  innertopmargin=8pt,
  innerbottommargin=8pt,
  skipabove=8pt,
  skipbelow=8pt
]{rqbox}

\newcommand{\RQbox}[2]{%
\begin{rqbox}
\textbf{#1}\par\medskip
\textit{#2}
\end{rqbox}
}

\newmdenv[
  backgroundcolor=green!8,
  leftline=true,
  rightline=false,
  topline=false,
  bottomline=false,
  linecolor=black,
  linewidth=3pt,
  innerleftmargin=10pt,
  innerrightmargin=10pt,
  innertopmargin=8pt,
  innerbottommargin=8pt,
  skipabove=8pt,
  skipbelow=8pt
]{resultbox}

\begin{document}

\title{Academic Integrity and Emotional Responses to Inappropriate LLM Use in Software Engineering Education}

\titlerunning{Academic Integrity and Emotional Responses to Inappropriate LLM Use}

\author{
Ronnie de Souza Santos\inst{1,2}
\and
Italo Santos\inst{3}
\and
Giuseppe Destefanis\inst{4}
\and
Cleyton Magalhães\inst{5}
\and
Mairieli Wessel\inst{6}
}

\authorrunning{R. de Souza Santos et al.}

\institute{
University of Calgary, Calgary, Canada
\and
CESAR School, Recife, Brazil
\and
University of Hawai‘i at Mānoa, USA
\and
University College London, United Kingdom
\and
Universidade Federal Rural de Pernambuco (UFRPE), Brazil
\and
Radboud University, The Netherlands
\\
\email{
ronniedesouzasantos@ucalgary.ca,
ress@cesar.school,
isantos3@hawaii.edu,
g.destefanis@ucl.ac.uk,
cleyton.vanut@ufrpe.br,
mairieli.wessel@ru.nl
}
}

\maketitle

\begin{abstract}

Academic integrity in higher education is increasingly shaped by complex socio-technical environments marked by automated tools, evolving institutional practices, and heightened performance pressures. Within this context, large language models (LLMs) are becoming prevalent in software engineering education, further blurring boundaries around acceptable assistance and authorship. This study investigates how software engineering students describe their emotional experiences after using LLMs in ways they perceive as academically inappropriate. We conducted a cross-sectional survey with 116 undergraduate students. Results show emotionally heterogeneous responses. Indifference was most frequent, including among students who recognized risks to learning and academic standing. Guilt and anxiety were reported in relation to moral discomfort and concern about penalties. Relief and satisfaction were evident primarily in deadline-driven contexts and situations of unclear guidance.


\end{abstract}

\section{Introduction}
\label{intro}

Academic integrity in higher education is commonly defined through adherence to explicit assessment rules and shared academic norms rather than as a single, uniformly interpreted behavior~\cite{whitley1998factors,mccabe2001cheating,sheard2002cheating}. Prior research characterizes cheating as a spectrum of practices, ranging from unauthorized collaboration and inappropriate code reuse to plagiarism, contract cheating, and examination misconduct~\cite{magnus2002tolerance,sheard2003investigating}. What constitutes cheating appears to be context-dependent, shaped by assessment design, disciplinary conventions, and the clarity of instructional guidance~\cite{hutton2006understanding,molnar2012does}. Empirical studies indicate that academic misconduct is widespread, with many students reporting engagement in at least one questionable practice during their academic trajectory~\cite{whitley1998factors,mccabe2001cheating}. Reported motivations include performance pressure, time constraints, perceived unfairness, peer normalization, and expectations regarding detection and institutional consequences~\cite{whitley1998factors,sheard2003investigating,magnus2002tolerance,mccabe2001cheating}.

Research addressing students' affective and moral experiences around cheating suggests that emotional responses are often complex and ambivalent. Prior to cheating, students frequently report stress and anxiety framed as situational pressures rather than moral motivations~\cite{hutton2006understanding,sheard2003investigating}. During decision making, moral neutralization is commonly reported, with students minimizing wrongdoing by emphasizing effort, intent, or perceived triviality~\cite{magnus2002tolerance,whitley1998factors}. After cheating, emotional responses vary, ranging from guilt and discomfort to limited emotional impact when behaviors are viewed as normative or low risk~\cite{sheard2002cheating,jones2011academic}. Stronger negative emotions are more frequently associated with examination-related misconduct, whereas lower-level practices, such as collaboration or reuse, tend to be accompanied by muted affect~\cite{molnar2012does,sheard2002cheating}.

Recent work indicates that large language models (LLMs) have altered the contemporary cheating landscape, particularly in computing and software engineering education. LLMs enable rapid generation of code and text, complicating notions of authorship and effort~\cite{cotton2024chatting,firth2024cheating,rios2023authorship}. Students frequently report uncertainty about acceptable boundaries, especially for debugging, code generation, and textual reformulation~\cite{lee2024cheating,mah2024beyond}. LLM-assisted cheating is often framed pragmatically, emphasizing efficiency, perceived norms, and low detectability, while technical and pedagogical research highlights increased opportunity for misconduct and challenges to detection~\cite{rytilahti2024easy,ortiz2025chat,alzahrani2022detecting,salim2024impeding}. Although this literature documents shifting practices and institutional responses, it offers limited insight into students' emotional experiences when they perceive LLM use as academically inappropriate~\cite{mah2024beyond,lee2024cheating,marlowe2026systemic,randall2023ai,santos2026llm}. In this sense, the present study focuses on the feelings of software engineering students, considering what they interpret as academically inappropriate use of LLM by answering the following research question (RQ):

\RQbox{RQ$_1$ -- Students’ Emotional Responses to Academically Inappropriate LLM Use}
{How do software engineering students describe their emotional experiences after using large language models in ways that violate academic integrity expectations?}

To answer our RQ, we conducted a cross-sectional survey examining software engineering students' self-reported experiences using LLMs in ways they perceived as academically inappropriate or misaligned with course expectations, combining closed-ended questions to characterize usage contexts with open-ended questions to capture students' reflections on feelings and perceived consequences following such use. Our findings show that students report emotionally heterogeneous experiences after academically inappropriate LLM use, most often marked by indifference. While guilt and anxiety occur, they are typically situational, and positive emotions such as relief sometimes arise under academic pressure. Overall, LLM misuse appears normalized, with awareness of integrity violations often coexisting with neutral or mixed emotional responses rather than strong distress.

The contribution of this study lies in its focus on software engineering students' affective experiences regarding academic integrity, an aspect that remains underexplored in software engineering education research. Prior work in software engineering education has primarily focused on defining misconduct, identifying vulnerable assessment formats, or proposing technical and policy-oriented responses to cheating, including in the context of LLM use~\cite{sheard2002cheating,rytilahti2024easy,salim2024impeding,cotton2024chatting,marlowe2026systemic}. Although recent studies acknowledge uncertainty and pragmatic reasoning around LLM-assisted work, students' emotional responses to engaging in academically inappropriate use of these tools have received limited attention~\cite{lee2024cheating,mah2024beyond}. This study foregrounds students' reported feelings, filling a gap in software engineering education research and offering an empirically grounded perspective on how academic integrity is experienced and interpreted within AI-mediated learning contexts.
\section{Background}
\label{sec:background}

This section reviews prior literature on cheating and academic misconduct in higher education, focusing on how practices, interpretations, and responses have evolved over time.

\subsection{Academic Integrity in Higher Education}

Academic integrity has evolved alongside changes in educational practices and technologies. Early discussions emerged in contexts where assessments were primarily paper based and conducted under direct supervision, limiting opportunities for misconduct and framing cheating largely as an individual moral failure associated with honesty, character, and personal responsibility~\cite{harp1965academic,newton2016academic}. Student experiences received little attention during this period, as misconduct was generally interpreted as a deliberate ethical lapse rather than a response to contextual factors~\cite{harp1965academic,macfarlane2014academic}. With the expansion of higher education and the introduction of mass assessment practices, research began documenting academic misconduct as a widespread phenomenon across institutions and disciplines~\cite{whitley1998factors,mccabe2001cheating}. Studies increasingly identified stress, performance pressure, peer influence, and perceptions of fairness as factors associated with cheating, while emotional responses such as guilt and discomfort were often moderated by normalization and perceptions of low risk~\cite{whitley1998factors,mccabe2001cheating,magnus2002tolerance}.

The widespread adoption of personal computers, internet access, and digital submission systems further transformed academic integrity landscapes by expanding opportunities for plagiarism, unauthorized collaboration, and access to external resources~\cite{macfarlane2014academic,newton2016academic}. These developments also increased ambiguity regarding acceptable academic practices, prompting researchers to emphasize the role of unclear expectations, inconsistent enforcement, and performance oriented educational cultures in shaping misconduct~\cite{macfarlane2014academic,simon2016negotiating,newton2016academic}. During this period, academic integrity came to be viewed less as a fixed standard and more as a negotiated and relational concept shaped by interactions among students, instructors, and institutional policies~\cite{simon2016negotiating}. Research also began to examine students' experiences more directly, documenting feelings of anxiety, confusion, fear, and perceived injustice in response to allegations, investigations, and sanctions~\cite{stone2023student,newton2016academic}.

More recently, academic integrity has been examined within complex socio technical systems characterized by automated tools, globalized higher education, and heightened performance pressures~\cite{sanni2021international,eaton2023academic}. Contemporary perspectives emphasize academic integrity as a systemic and educational concern that depends on institutional communication, transparency, consistency, and student engagement~\cite{macfarlane2014academic,eaton2023academic}. The emergence of LLMs represents the latest development in this trajectory, introducing tools capable of generating text, code, and solutions while blurring boundaries around authorship, originality, and acceptable assistance~\cite{cotton2024chatting,firth2024cheating,randall2023ai,santos2026llm}. Recent studies suggest that students' experiences in AI mediated contexts are shaped not only by rule interpretation but also by emotional responses such as uncertainty, discomfort, resignation, and normalization, which are intertwined with perceptions of fairness, effort, and peer norms~\cite{stone2023student,lee2024cheating,mah2024beyond,ortiz2025chat}. Consequently, academic integrity is increasingly understood as an ongoing process of sense making shaped by evolving technologies, institutional responses, and students' lived experiences~\cite{eaton2023academic,stone2023student}.

\subsection{Academic Integrity in Software Engineering Education}

Research on academic integrity in software engineering education largely reflects patterns identified in the broader higher education literature, rather than proposing a distinct disciplinary conceptualization of cheating. Existing studies generally adopt prevailing definitions of academic misconduct and apply them to software engineering contexts, particularly courses centered on programming intensive assessments~\cite{sheard2002cheating}. Within this body of work, academic integrity is primarily framed in terms of compliance with assessment rules and submission requirements, with limited attention to professional ethics or affective experience in early research~\cite{sheard2002cheating}.

A defining characteristic of this literature is its narrow emphasis on specific forms of coursework. Most studies focus on individual coding assignments and examinations, where misconduct is discussed in relation to unauthorized code reuse, plagiarism, and inappropriate collaboration~\cite{sheard2002cheating,sheard2003investigating}. Other core software engineering activities, including requirements analysis, architectural design, testing, and maintenance, receive comparatively little attention, reflecting a focus on assessable artifacts rather than process oriented work. Within programming focused contexts, students are shown to differentiate among forms of misconduct, often viewing collaboration, partial reuse, or reference to external examples as more acceptable than direct copying of complete solutions~\cite{sheard2002cheating,sheard2003investigating}. These interpretations vary across assessment structures and instructional settings, indicating that academic integrity judgments are shaped by task design and guidance rather than uniform adherence to institutional policy~\cite{sheard2003investigating}.

Recent research addressing LLMs extends established patterns rather than redefining academic integrity in software engineering education. Empirical studies demonstrate that LLMs can generate correct or near correct solutions for a substantial portion of programming assignments, challenging assumptions about individual authorship and effort~\cite{rytilahti2024easy}. Other work shows that LLM assisted cheating is technically feasible while also suggesting that assessment structure and design may partially constrain such use~\cite{salim2024impeding}. Students frequently report uncertainty about acceptable boundaries when using LLMs, emphasizing pragmatic considerations such as efficiency, perceived norms, and alignment with course expectations~\cite{lee2024cheating,mah2024beyond,ortiz2025chat}. Conceptual analyses argue that these challenges cannot be addressed through detection or enforcement alone, framing academic integrity as a negotiated process shaped by tools, assessments, and institutional expectations~\cite{randall2023ai,cotton2024chatting,firth2024cheating}. Despite growing attention to LLM related misconduct, empirical work examining how integrity is taught and emotionally experienced across the full range of software engineering activities remains limited~\cite{lee2024cheating,mah2024beyond}.

\section{Method}
\label{sec:method}

In this study, we examined software engineering students' self-reported experiences using LLMs in ways they perceived as inappropriate, disallowed, or misaligned with course expectations. Following prior research on academic misconduct and LLM assisted cheating~\cite{sheard2002cheating,sheard2003investigating,hosny2014attitude,lee2024cheating,cotton2024chatting,adnan2025cheating,ortiz2025chat}, we employed a cross-sectional survey design~\cite{pfleeger2001principles,easterbrook2008selecting,ralph2020empirical}. The methodological emphasis of this study was students' emotional experiences following LLM use as they associated it with academic integrity violations. Our survey therefore combined closed-ended questions, used to characterize contexts and patterns of use, with open-ended questions designed to elicit reflective accounts of feelings, interpretations, and perceived consequences.

\paragraph{\textbf{Survey Design.}}
In line with established guidelines for survey based software engineering research~\cite{pfleeger2001principles,linaker2015guidelines}, we designed an anonymous online survey to capture students' accounts of LLM use that conflicted, in their view, with course rules, assessment expectations, or academic integrity policies. The survey was implemented using Qualtrics\footnote{[www.qualtrics.com](http://www.qualtrics.com)} and organized into five thematic sections, followed by demographic and contextual questions. The instrument was informed by prior empirical research on academic integrity, student cheating, and the use of LLMs in educational contexts~\cite{mccabe2001cheating,sheard2002cheating,sheard2003investigating,molnar2012does,cotton2024chatting,lee2024cheating,ortiz2025chat}. Building on these foundations, the survey was constructed to prompt participants to describe concrete situations in which LLM use was perceived as academically inappropriate, rather than to elicit abstract opinions or hypothetical judgments. Participants were asked to report the types of coursework involved, the forms of assistance used, and the situational conditions surrounding these decisions. In addition to motivations, constraints, and justifications, particular attention was given to emotional responses reported after such use. The survey also captured participants' perceptions of where within software engineering curricula these practices were most prevalent, together with their views on academic, personal, and professional consequences. Closed-ended items supported the identification of recurring patterns related to the theme, while open-ended questions enabled detailed descriptions and sense making accounts. Background data were collected for context, and no personally identifiable information was requested.

\paragraph{\textbf{Pilot.}}
After initial development and internal review, the survey instrument underwent a pilot phase to refine both technical functionality and content clarity. The technical pilot involved testing the survey across multiple browsers and devices to confirm correct display, navigation flow, and response recording. A content-focused pilot was subsequently conducted with two software engineering faculty members and three undergraduate students enrolled at different stages of a software engineering program. Pilot participants completed the survey and provided feedback on question clarity, interpretability, and relevance. Their feedback informed revisions to wording, sequencing, and response options, with particular attention to ensuring that questions encouraged reflection on personal experiences and emotional reactions rather than generalized statements or normative judgments. The complete set of survey questions analyzed for this study is presented in Table~\ref{tab:surveyquestions}.

\small

\begin{longtable}{p{0.5cm}|p{11.0cm}}
\caption{Survey Questionnaire}
\label{tab:surveyquestions}\\
\hline

\multicolumn{2}{l}{\textbf{EXPERIENCES WITH INAPPROPRIATE LLM USE}} \\
\hline \hline

1 &
Describe a situation in your software engineering program where you used an LLM in a way that you believe was not aligned with the course rules or expectations.
\\
\hline

2 &
In what types of coursework have you used an LLM in ways that were likely not permitted? (Select all that apply)

\vspace{0.1cm}

{\scriptsize
A. ( ) Regular classwork or weekly exercises |
B. ( ) Programming assignments or labs |
C. ( ) Group or capstone projects |
D. ( ) Essays or written reflections |
E. ( ) Software design or documentation tasks |
F. ( ) Quizzes or short online tests
G. ( ) Major exams or finals |
H. ( ) Other
}
\\
\hline

3 &
Which of the following LLM uses have you personally engaged in, even if they were not allowed by the course? (Select all that apply)

\vspace{0.1cm}

{\scriptsize
A. ( ) Submitting LLM-generated code or solutions as my own work |
B. ( ) Copying LLM-generated text into reports or essays without acknowledgment |
C. ( ) Using LLMs during quizzes, tests, or exams |
D. ( ) Having LLMs perform debugging or test-case generation for assessed work |
E. ( ) Asking LLMs to produce substantial parts of group project deliverables |
F. ( ) Using multiple LLMs or rewriting strategies to obscure AI involvement |
G. ( ) Other use that conflicted with course expectations
}
\\
\hline \hline

\multicolumn{2}{l}{\textbf{MOTIVATIONS AND DECISION FACTORS}} \\
\hline \hline

4 &
What were your main reasons for using an LLM in this way?
\\
\hline

5 &
How did you interpret or justify this decision at the time?

\vspace{0.1cm}

{\scriptsize
A. ( ) I felt it was necessary to manage the workload |
B. ( ) I believed the rules were unclear or outdated |
C. ( ) I thought the impact was minor |
D. ( ) I avoided thinking about possible consequences |
E. ( ) I knew it violated expectations but accepted the risk
}
\\
\hline

6 &
Afterward, which emotion best describes how you felt?

\vspace{0.1cm}

{\scriptsize
A. ( ) Relief |
B. ( ) Guilt |
C. ( ) Indifference |
D. ( ) Satisfaction |
E. ( ) Anxiety
}
\\
\hline

\multicolumn{2}{l}{\textbf{AFFECTED AREAS OF THE PROGRAM}} \\
\hline \hline

7 &
In your experience, which aspects of software engineering education are most affected by inappropriate or unclear LLM use? (Select all that apply)

\vspace{0.1cm}

{\scriptsize
A. ( ) Programming and implementation |
B. ( ) Software testing or debugging |
C. ( ) System design or architecture |
D. ( ) Project management and teamwork |
E. ( ) Technical documentation or reports |
F. ( ) Individual skill assessments
}
\\
\hline

8 &
Which course contexts made it easiest to rely on LLMs in ways that conflicted with expectations?

\vspace{0.1cm}

{\scriptsize
A. ( ) Introductory programming courses |
B. ( ) Advanced software engineering or design courses |
C. ( ) Testing and quality assurance courses |
D. ( ) Capstone or group projects |
E. ( ) Theory or writing-focused courses such as ethics or requirements
}
\\
\hline \hline

\multicolumn{2}{l}{\textbf{PERCEIVED CONSEQUENCES AND DETERRENTS}} \\
\hline \hline

9 &
What personal, academic, or professional consequences do you think may result from relying on LLMs in ways that bypass learning or assessment goals?
\\
\hline \hline

\multicolumn{2}{l}{\textbf{DEMOGRAPHIC}} \\
\hline \hline

10 &
Year in your program?
\\
\hline

11 &
Have you received any formal training or guidance on responsible LLM use in your program?

\vspace{0.1cm}

{\scriptsize
A. ( ) Yes |
B. ( ) No |
C. ( ) Not sure
}
\\
\hline

12 &
In the courses where this occurred, were expectations about LLM use communicated?

\vspace{0.1cm}

{\scriptsize
A. ( ) Yes, clearly and explicitly |
B. ( ) Yes, but only in general terms |
C. ( ) No guidance was provided |
D. ( ) I do not remember
}
\\
\hline

13 &
What best describes your gender identity?
\\
\hline

14 &
Country where you are currently studying?
\\
\hline \hline

\end{longtable}

\paragraph{\textbf{Recruitment.}}
Participant recruitment followed recommendations for empirical software engineering studies and combined platform based sampling with referral based snowball sampling~\cite{ralph2020empirical,baltes2022sampling}. The primary recruitment channel was the Prolific platform, which supports structured prescreening and has been widely used in survey research within software engineering~\cite{russo2022recruiting,reid2022software}. Platform filters were configured to identify individuals who reported current enrollment in undergraduate software engineering and prior experience using LLMs in academic coursework. These criteria were supplemented by eligibility questions embedded within the survey to confirm participants' enrollment status and LLM experience. To broaden participation beyond the Prolific pool, respondents were invited to share the study with peers enrolled in similar programs who met the same eligibility requirements, following a snowball sampling approach~\cite{baltes2022sampling}. Data collection remained open for two weeks and initially yielded 208 responses from software engineering students across different stages of their programs and institutional contexts. After applying eligibility verification, attention checks, and quality screening procedures, 60 responses were removed, resulting in 148 responses. Subsequently, a manual review identified 32 responses with indications of AI generated or low authenticity content, resulting in a final dataset of 116 responses.

\paragraph{\textbf{Filtering.}}
Following data collection, responses were subjected to a multi-stage filtering process to ensure data validity and quality, in accordance with recommendations for survey based software engineering research using online platforms~\cite{danilova2021you,alami2024you}. First, submissions were screened against eligibility criteria using both platform level prescreening attributes and survey level validation questions. Incomplete responses and submissions that failed these checks were removed. Second, attention and engagement checks were applied. Responses that failed attention checks, showed uniform answering patterns across closed-ended items, or exhibited unusually short completion times were excluded, as these patterns suggested limited engagement. Third, given the study's focus on reflective accounts of personal experience, responses were reviewed for indications of automated or AI generated text. Submissions characterized by generic phrasing, repetitive structure, or a lack of situational specificity were excluded. This step aimed to ensure that the dataset reflected participants' own experiences and emotional interpretations rather than generated content.

\paragraph{\textbf{Data Analysis.}}
The survey produced both qualitative and quantitative data. Closed-ended questions supported descriptive quantitative analysis, while open-ended questions examining experiences, motivations, emotions, and perceived consequences supported the qualitative coding process. Quantitative survey items were analyzed using descriptive statistics to summarize reported LLM usage, coursework contexts, prior experience, and perceived risks~\cite{george2018descriptive}. Frequency distributions and summary measures were used to characterize participant backgrounds and situational factors and to support interpretation of the qualitative findings, rather than to enable inferential claims.

Open-ended responses were analyzed using an iterative inductive qualitative coding and categorization process informed by established qualitative guidelines in software engineering research~\cite{cruzes2011recommended}. Responses were read in full to identify recurring experiences, motivations, emotional reactions, and contextual factors associated with perceived inappropriate LLM use. Initial inductive codes captured recurring patterns across participants' narratives and were subsequently grouped into broader interpretive categories aligned with the research questions. Categories included contextual pressures, interpretations of course expectations, emotional reactions, perceived learning impacts, and anticipated academic consequences. Coding was conducted by at least two researchers, with discrepancies resolved through discussion and consensus. Table~\ref{tab:codingexample} presents an example of the coding and categorization process used during the analysis.

\begin{table*}[t]
\caption{Example of Qualitative Coding and Categorization}
\label{tab:codingexample}
\centering
\small
\begin{tabular}{|p{5.0cm}|p{2.5cm}|p{2.5cm}|p{2.8cm}|}
\hline

\textbf{Participant Response}
&
\textbf{Core Idea}
&
\textbf{Low-Level Code}
&
\textbf{Category}
\\
\hline

I used ChatGPT to do coding and sometimes to find information about a research topic.
&
``used ChatGPT to do coding''
&
Coding support
&
\multirow{3}{=}{Coding and Development Support}
\\
\cline{1-3}

I use LLMs especially for things like generation of the initial structure (boilerplate) in order to save time. I input existing functions to find efficient ways to structure the logic.
&
``generation of the initial structure''
&
Boilerplate generation
&
\\
\cline{1-3}

Writing code and fixing bugs and testing.
&
``writing code and fixing bugs and testing''
&
Software development support
&
\\
\hline

\end{tabular}
\end{table*}

To support characterization of affective responses, the qualitative categories were subsequently interpreted using the Self Assessment Manikin (SAM) framework~\cite{bradley1994measuring}, focusing on valence and arousal dimensions. Valence was identified from participants' emotional descriptions and evaluative language, while arousal was inferred from contextual cues such as urgency, stress, fear of failure, or calm acceptance. This interpretive use of the SAM framework allowed affective responses to be characterized systematically while remaining grounded in participants' own accounts.

\paragraph{Ethics.}
The study was conducted in accordance with institutional guidelines for research involving human participants and received approval from the first author's university ethics board. Informed consent was obtained prior to participation, and respondents were informed about the study's purpose, voluntary nature, data handling procedures, and their right to withdraw at any time. No personally identifiable information was collected, and findings are reported in aggregate form. These measures were particularly important given the study's focus on experiences involving potentially disallowed academic practices, where confidentiality and participant protection were necessary to support ethical data collection and candid reporting.

\section{Results}
\label{sec:findings}

In this section, we report the survey results, starting with participant demographics, LLM usage and perceptions of cheating, students’ emotional responses, and how these relate to demographic and usage contexts, supported by illustrative participant quotations.

\subsection {Participant Demographics}

The survey comprises responses from 116 undergraduate students enrolled in software engineering programs. Participants were distributed across different stages of their undergraduate studies, with 19.8\% (n=23) in their first year, 16.4\% (n=19) in their second year, 28.4\% (n=33) in their third year, and 35.3\% (n=41) in their fourth or fifth year. This distribution indicates that a substantial portion of the sample consisted of students in later stages of their programs. Participants also reported varied institutional contexts regarding guidance on LLM use. More than half of the respondents, 54.3\% (n=63), indicated that they had not received formal training or guidance on responsible LLM use within their program, while 43.1\% (n=50) reported having received some form of formal guidance. A small proportion, 2.6\% (n=3), were unsure. At the course level, expectations regarding LLM use were most commonly communicated in general terms, as reported by 53.4\% (n=62). Clear and explicit communication was reported by 23.3\% (n=27), whereas 18.1\% (n=21) indicated that no guidance was provided and 5.2\% (n=6) did not recall whether expectations had been communicated. In addition, 53.4\% (n=62) reported that their institution had a formal policy addressing AI or LLM use in coursework, while 42.2\% (n=49) were not aware of such a policy and 4.3\% (n=5) were unsure. 

In terms of geographic distribution, participants were studying in a wide range of countries. The largest groups were based in South Africa (19.8\% – n=23), the United Kingdom (10.3\% – n=12), the United States of America (9.5\% – n=11), Canada (8.6\% – n=10), and Germany (6.9\% – n=8). Additional participants reported studying in Italy (5.2\% – n=6), Hungary and India (4.3\% – n=5 each), France, Mexico, the Philippines, Poland, and Portugal (2.6\% – n=3 each), and Brazil, Chile, Greece, and Latvia (1.7\% – n=2 each). Smaller representations were observed from Australia, Austria, the Czech Republic, Egypt, Ireland, Kenya, the Netherlands, New Zealand, Sweden, and Vietnam (0.9\% – n=1 each). Finally, with respect to gender identity, 62.9\% (n=73) identified as men, 36.2\% (n=42) identified as women, and 0.9\% (n=1) identified as non-binary. Gender identity was self-reported, and no additional categories were selected.

\subsection{Software Engineering Course Contexts Associated with Inappropriate LLM Use}

A total of 52 students (44.8\%) reported \textbf{introductory programming courses} as contexts in which they frequently relied on LLMs in ways that conflicted with course expectations. These accounts were often described around limited prior knowledge, rapid learning pace, and time pressure early in the curriculum. One participant described turning to LLMs when struggling to begin programming tasks: \textit{``In the first year of my studies I used AI tools quite often, usually for coding homework. It was because I didn't understand anything about programming, I was lost and didn't know where to even begin''} (P084). Others emphasized deadline proximity, for example, \textit{``it was a python assignment and i only had 4 hours left so i had to use a LLM''} (P011). These narratives suggest that introductory courses were perceived as particularly vulnerable due to a combination of skill gaps and workload pressure.

A similarly high proportion of participants, 52 students (44.8\%), identified \textbf{theory- or writing-focused courses}, such as ethics or requirements engineering, as contexts where inappropriate LLM use was necessary. Students described using LLMs to structure written work, generate text, or validate their ideas, often citing insecurity or low engagement with the subject matter. One participant explained, \textit{``I use LLM for tasks where I struggle to organize my ideas… such as in reports''} (P021). Another reported delegating most of the writing to an LLM due to lack of interest: \textit{``I used an LLM for my assignment where I had to write an essay on digital leadership… and asked the LLM to write the assignment for me''} (P085). These responses indicate that writing-intensive contexts were perceived as amenable to LLM substitution rather than assistance.

\textbf{Advanced software engineering or design courses} were reported by 46 participants (39.7\%). In these contexts, students described conceptual complexity, unfamiliar notations, or advanced technical requirements as drivers of LLM use. One participant noted relying on an LLM to navigate course concepts despite explicit prohibitions: \textit{``In my most recent class in Object Orient Analysis and design… I used an LLM to help me understand what in the world was going on everything from the diagrams to understanding concepts''} (P013). Another described submitting substantial code fragments to resolve concurrency issues under time pressure: \textit{``I put the whole worker\_thread function… into chatgpt which solved it''} (P072). These accounts suggest that advanced courses introduced cognitive and technical demands that shaped students’ reliance on LLMs.

\textbf{Testing and quality assurance courses} were selected by 35 participants (30.2\%). Students reported LLM use in these contexts as support for debugging, error detection, or test-related tasks, particularly when manual efforts had failed. One participant stated, \textit{``I had to quickly make a few bug fixes, I didn't have time to manually debug and I had already tried''} (P008). Another reflected on bypassing personal verification to save time: \textit{``I used an LLM to fix errors in my code without doublechecking the code myself''} (P039). These responses indicate that testing activities were perceived as areas where LLM assistance could easily extend into inappropriate reliance.

Finally, \textbf{capstone or group projects} were reported by 33 participants (28.4\%). Students described these settings as characterized by collective deadlines, uneven workload distribution, and high stakes, which shaped decisions to use LLMs beyond what was permitted. One participant recounted, \textit{``In the end I copied and pasted our work into ChatGPT and asked it to find the bugs. We were desperate at this point as the project was due in that week''} (P033). Another similarly noted, \textit{``I used an LLM to help me write the majority of the code… because i was the only one doing work so i neeeded help to finish in time''} (P113). These narratives indicate that collaborative project contexts introduced additional pressures that influenced LLM use.

\subsection{Perceived Consequences of Inappropriate LLM Use}
Overall, participants associated inappropriate LLM use with both learning-related and formal academic consequences. The most frequently affected areas were \textbf{programming and implementation} (80 participants, 69\%) and \textbf{software testing and debugging} (63 participants, 54.3\%), where students described reduced hands-on practice, weaker problem-solving abilities, and difficulties diagnosing errors when LLMs replaced rather than supported learning activities. Participants also identified impacts on \textbf{technical documentation and report writing} (55 participants, 47.4\%), \textbf{individual skill assessments} (54 participants, 46.6\%), \textbf{system design or architecture} (38 participants, 32.8\%), and \textbf{project management and teamwork} (37 participants, 31.9\%), suggesting concerns extending beyond implementation tasks to broader professional and collaborative competencies.

Participants also anticipated formal institutional consequences if inappropriate LLM use were detected. The most frequently reported outcome was receiving a zero grade or failing the course (52 participants, 44.8\%), followed by formal warnings or probation (30 participants, 25.9\%), and more severe sanctions such as suspension or expulsion (18 participants, 15.5\%). At the same time, \textbf{reduced skill development} and \textbf{overreliance on AI tools} were each reported by 99 participants (85.3\%), indicating widespread recognition that repeated misuse could affect long-term learning and preparedness for professional practice.

\subsection{Emotional Responses to Perceived Consequences of LLM Misuse}

Considering the emotional characteristics experienced by software engineering students when engaging with misuse or cheating using LLMs, five dominant sentiments emerged. \textbf{Indifference} was the most frequently reported sentiment, expressed by 42 participants (36.2\%). Participants expressing indifference often acknowledged potential consequences but described them in a detached, minimized, or pragmatic manner. In some cases, this involved dismissals such as \textit{``Mostly nothing happens'' (P018)} or conditional rationalizations such as \textit{``Professionally, none, so long as I can understand what is being output and can verify it. Academically, it could be more serious'' (P041}. Following this, \textbf{Guilt} was reported by 24 participants (20.7\%). Guilt captures moral discomfort and self-reproach associated with perceived overreliance on LLMs. Participants describing guilt frequently reflected on having undermined their own learning, confidence, or skill development. One participant stated: \textit{``Personally, it makes me a weaker programmer because I haven't spent enough time working hands-on with code.'' (P039)}. \textbf{Anxiety} about consequences was reported by 21 participants (18.1\%). This sentiment reflects concern about sanctions, academic failure, or long-term repercussions. Participants expressing anxiety frequently referenced institutional penalties, fear of detection, or the possibility of being unprepared for future assessments or professional work. One participant stated: \textit{``Chance of getting caught or flagged with the use of AI, resulting in losing the degree or having to redo courses.'' (P095)}. Also reported by 21 participants (18.1\%), \textbf{Relief} reflects reassurance or perceived mitigation of consequences. Participants reporting relief often qualified risks based on context, personal competence, or institutional ambiguity. One participant stated, \textit{``I do not think there are personal or academic consequences for me. I still complete my tasks and learn what is needed for the course.'' (P016)}. Finally, \textbf{Satisfaction} was the least reported sentiment, expressed by 8 participants (6.9\%). Satisfaction reflected positive evaluations of LLM use despite acknowledged risks. Participants expressing satisfaction emphasized perceived benefits such as efficiency or technological progress. For example, one participant described LLMs as \textit{``A huge improvement in tech'' (P092)}, while another stated, \textit{``Using LLMs can significantly boost the learning ability of students'' (P047)}.

Looking closely at the reported feelings and the narratives describing perceived consequences, we characterized these sentiments using valence and arousal as affective dimensions. \textbf{Indifference} appears to be primarily associated with \textbf{neutral valence} and spans a \textbf{low to high range of arousal}. Participants reporting indifference often describe potential academic or professional consequences, including failure, expulsion, or long-term skill erosion, in a detached or matter-of-fact manner. In lower arousal cases, indifference may indicate normalization or minimization of perceived consequences. When combined with moderate or higher arousal, it suggests that participants recognize the presence of risk while not experiencing it as emotionally distressing. \textbf{Guilt} is associated with \textbf{negative valence} and is more commonly paired with \textbf{moderate arousal}. Participants reporting guilt tend to frame consequences in terms of moral discomfort, self-blame, or concerns about having undermined their own learning. Compared to anxiety, these narratives less frequently emphasize immediate punishment or urgency, and instead focus on internal evaluation. \textbf{Anxiety about consequences} is characterized by \textbf{negative valence} combined with \textbf{moderate to high arousal}. Participants reporting anxiety often reference fear of sanctions, academic failure, or long-term repercussions. The associated arousal reflects heightened concern in relation to perceived threat or uncertainty. In contrast to guilt, anxiety appears more strongly oriented toward external or institutional outcomes. \textbf{Relief} is associated with \textbf{positive valence}, but is often accompanied by \textbf{moderate to high arousal}. Participants expressing relief frequently describe pressure-filled contexts, such as heavy workloads or unclear expectations, alongside narratives that emphasize perceived mitigation of consequences. In this sense, relief appears to reflect a reduction in perceived stress rather than an absence of concern. \textbf{Satisfaction} is associated with \textbf{positive valence} and is most often paired with \textbf{low arousal}. Participants expressing satisfaction tend to emphasize instrumental aspects, such as efficiency or perceived benefit. These narratives rarely convey urgency or threat, suggesting that satisfaction reflects pragmatic approval rather than heightened emotional engagement.

Overall, the valence–arousal distribution observed in the data points to a possible dissociation between perceived consequences and emotional engagement. While many participants articulated serious academic or professional risks, these perceptions were not uniformly accompanied by negative affect or heightened emotional distress. Instances of neutral valence paired with moderate or high arousal were frequently observed, suggesting that recognition of potential consequences can occur alongside a relatively detached affective stance of the misuse. These observations indicate that students’ emotional responses to LLM misuse and cheating practices may reflect not only awareness of potential harm, but also broader contextual factors such as normalization of risk, institutional expectations, or perceived necessity. In this sense, awareness of consequences alone may not be sufficient to consistently produce strong affective responses.

\subsection{Relations Among Academic Stage, Consequences, and Emotional Responses}

\begin{figure*}[!bth]
    \centering
    \includegraphics[width=1\textwidth]{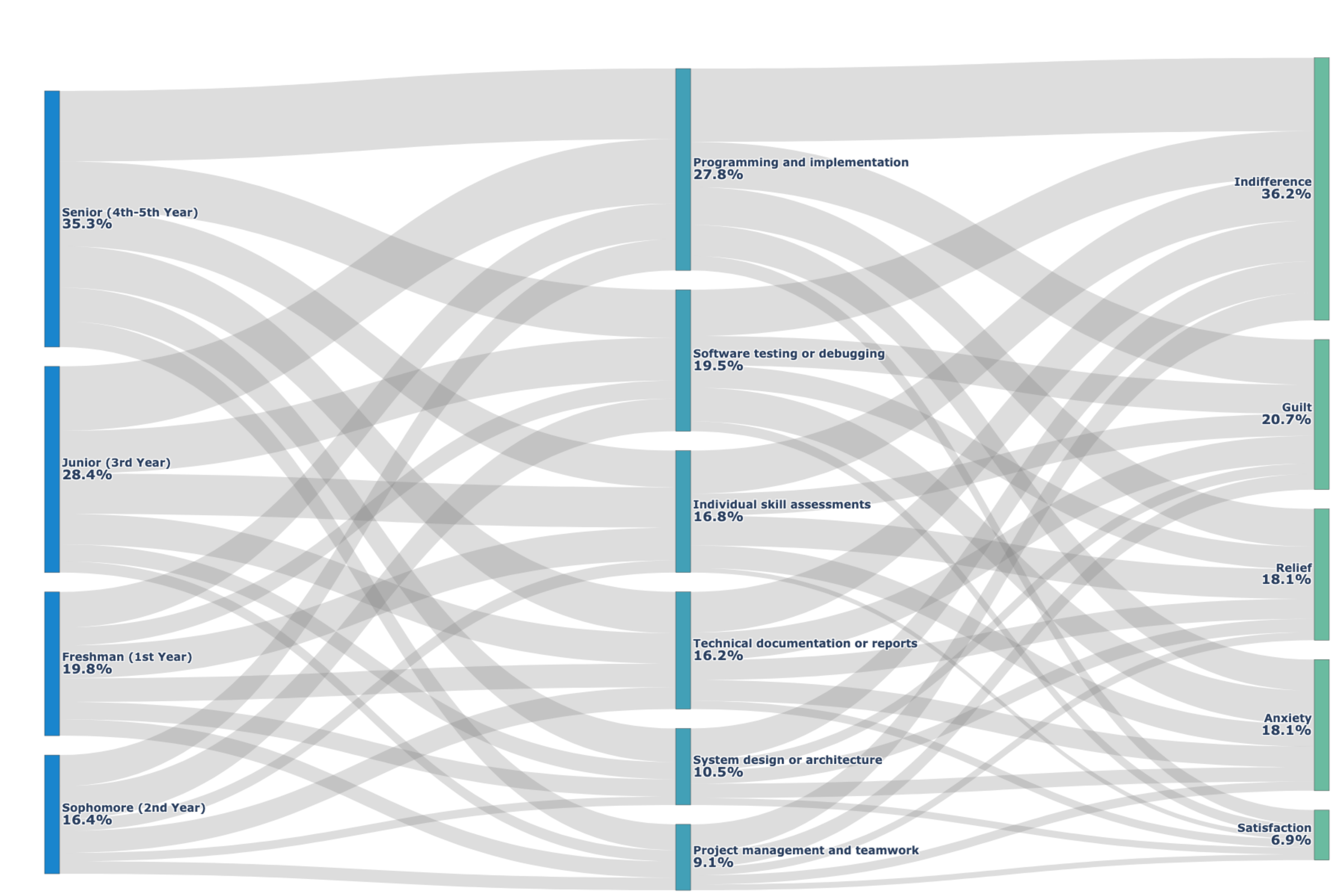}
    \caption{Weighted Sankey diagram showing co-occurrence relationships among reported emotional responses, academic stage, and software engineering topics associated with perceived inappropriate LLM use. Node and link widths are proportional to respondent weights and do not represent mutually exclusive transitions.}
    \label{fig:sankey_topics}
    \vspace{-10px}
\end{figure*}

\begin{figure*}[!bth]
    \centering
    \includegraphics[width=1\textwidth]{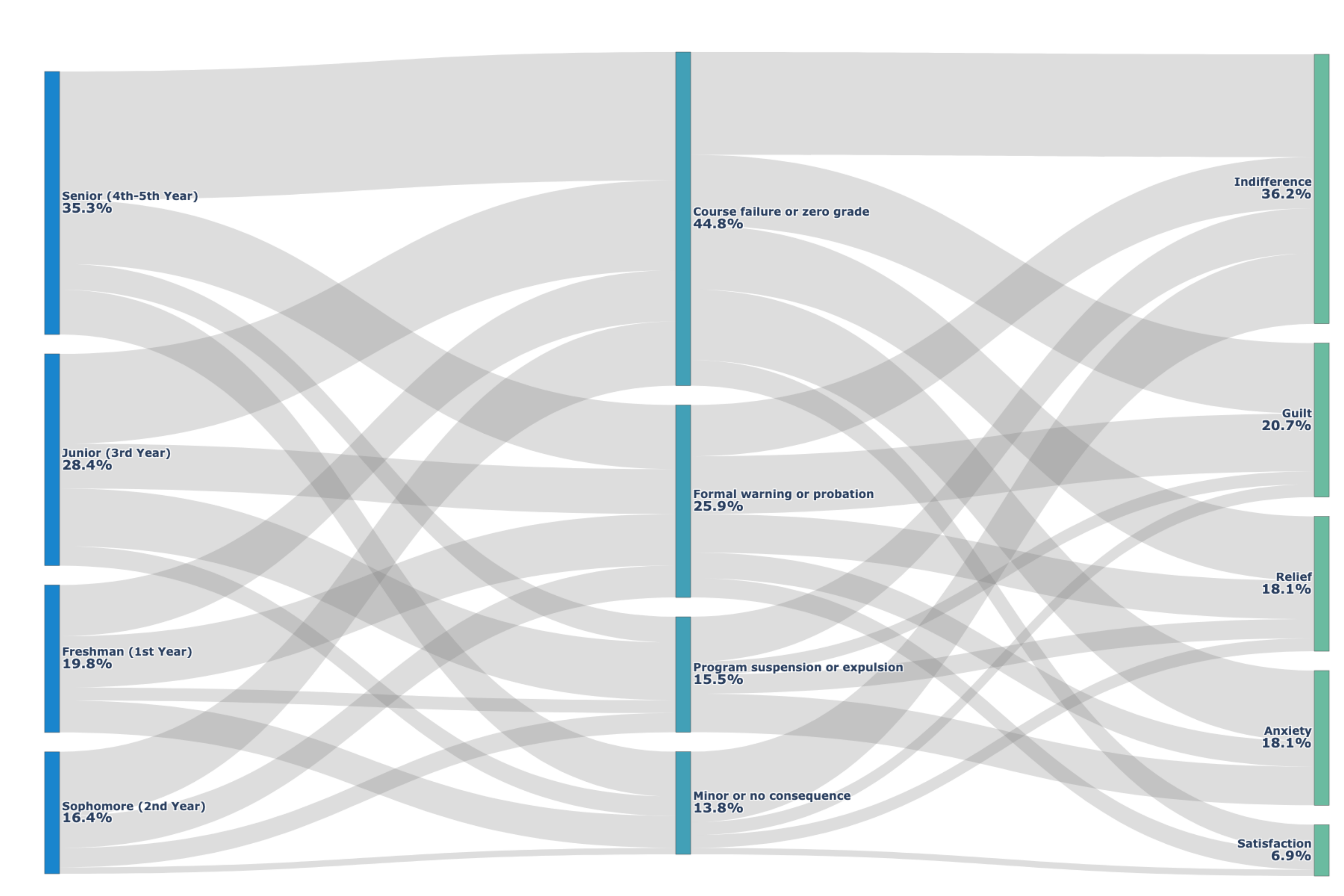}
    \caption{Weighted Sankey diagram showing co-occurrence relationships among reported emotional responses, academic stage, and anticipated academic consequences of perceived inappropriate LLM use. Node and link widths are proportional to respondent weights and do not represent mutually exclusive transitions.}
    \label{fig:sankey_consequences}
    \vspace{-10px}
\end{figure*}

Figures~\ref{fig:sankey_topics} and~\ref{fig:sankey_consequences} show how reported emotional responses are distributed across academic stage (e.g., freshman, sophomore, junior and senior), software engineering topics, and anticipated academic consequences of perceived inappropriate LLM use. Emotional responses appear across all years of study, with no clear concentration in a specific academic stage. Indifference is present among students at every stage and is connected to a wide range of contexts and consequences, indicating that emotionally neutral responses to perceived misuse are not restricted to early exposure or limited experience within the curriculum.

Figure~\ref{fig:sankey_topics} relates academic stage, software engineering topics, and emotions. Core curricular areas, including programming and implementation, software testing or debugging, and individual skill assessments, are associated with multiple emotional responses rather than a single dominant affective stance. Indifference and guilt appear most frequently across these topics, while anxiety and relief are also reported in technically demanding or assessment-focused contexts. More advanced areas, such as system design or architecture, and project management or teamwork display similar emotional distributions, although with lower overall frequency, suggesting that emotional responses are not topic specific but recur across different forms of technical engagement.

Figure~\ref{fig:sankey_consequences} connects the academic stage, anticipated academic consequences, and emotional responses. Severe institutional outcomes, including course failure, probation, and program suspension, are linked to both negative and neutral emotional responses. Indifference remains prominent even when participants report awareness of high stakes consequences, while guilt and anxiety appear alongside both severe and minor anticipated outcomes. Relief is also observed across consequence categories, particularly in contexts where participants describe situational pressure or uncertainty. These distributions show that recognition of serious academic consequences does not consistently correspond to heightened emotional engagement.


\begin{resultbox}
\faChartBar\ \textbf{RQ$_1$ Result: Emotional Experiences After Academically Inappropriate LLM Use}\\

Students describe emotionally heterogeneous experiences, most often marked by limited distress.

\medskip
\textbf{Key findings}

\begin{itemize}
\item \textbf{Indifference} is the most common response, even when students recognize risks to learning.
\item \textbf{Negative emotions} such as guilt and anxiety appear but are typically situational.
\item \textbf{Positive emotions}, including relief and occasional satisfaction, arise mainly under time pressure, technical difficulty, or unclear expectations.
\end{itemize}

\medskip
Overall, students describe their emotional experiences as shaped by routine academic pressures and normalized use of LLMs, where awareness of academic integrity violations coexists with emotionally neutral or mixed responses rather than consistently heightened emotional engagement.
\end{resultbox}

\section{Discussion} 
\label{sec:discussion}

This section situates the findings within prior academic integrity research and software engineering education literature. It discusses how the results relate to existing knowledge, outlines implications for research and educational practice, and considers threats to validity.

\subsection{Comparing Findings and Novelty}

Our results align with prior academic integrity research in higher education showing that cheating is shaped by contextual pressures, interpretive ambiguity, and students' situated judgments rather than by isolated moral failure~\cite{whitley1998factors,magnus2002tolerance,macfarlane2014academic,newton2016academic}. Across this literature, students are shown to distinguish between practices framed as clearly prohibited and those perceived as negotiable, particularly in contexts marked by workload pressure, ambiguous guidance, or low perceived oversight~\cite{whitley1998factors,magnus2002tolerance,macfarlane2014academic,newton2016academic}. Our study situates these established patterns within contemporary software engineering education, where students describe LLM-assisted cheating as embedded in routine coursework rather than as deliberate transgression. In our results, this negotiation is especially visible in software engineering contexts such as programming assignments, labs, and documentation tasks, where recurring deadlines and cumulative effort shape judgments about acceptable LLM use.

Prior academic integrity research on plagiarism, online resources, and contract cheating indicates that students adapt misconduct practices to available tools while relying on recurring justifications related to efficiency, fairness, and perceived necessity~\cite{whitley1998factors,magnus2002tolerance,jones2011academic,newton2016academic,sanni2021international}. Our study situates this continuity within software engineering education, where LLMs intersect with artifact-based assessment and professional norms that already value reuse and automation. Our results show that students extend these justifications beyond programming to software engineering activities such as debugging, testing, system design, and group projects, indicating that LLM-assisted cheating affects a wider set of learning processes than those emphasized in earlier integrity research focused primarily on programming artifacts~\cite{sheard2002cheating,sheard2003investigating}.

Academic integrity literature has increasingly acknowledged affective dimensions of misconduct, often reporting guilt, anxiety, or fear in relation to detection and sanctions~\cite{newton2016academic,stone2023student}. However, within this body of work, emotional accounts are typically treated as secondary to behavioral descriptions and are infrequently addressed in a systematic manner~\cite{macfarlane2014academic,newton2016academic,stone2023student}. Our study situates emotional response as a central analytic dimension within software engineering education. Our results show that indifference is the most frequently reported emotional response following perceived LLM misuse, even when students articulate substantial learning-related and academic risks. In our results, this emotional neutrality coexists with explicit recognition of learning loss and academic risk within software engineering education, suggesting that emotional response and consequence awareness are only loosely connected in this context.

While prior work documents uncertainty and ambiguity around AI use in academic settings~\cite{cotton2024chatting,lee2024cheating,mah2024beyond,ortiz2025chat}, our results provide empirical evidence from software engineering education that emotionally muted responses can accompany clear awareness of educational and professional consequences. The novelty of our study lies in integrating emotional characterization with contextualized accounts of LLM-assisted cheating across a range of software engineering activities. Grounded in students' reported experiences and situated within established academic integrity research, our findings show that LLM-assisted cheating in software engineering is a normalized and affectively heterogeneous practice, sustained by routine academic pressures and negotiated interpretations of acceptable use rather than by lack of awareness of its consequences.

\subsection{Implications}

Our results suggest that research on academic integrity and LLM use should move beyond documenting prevalence or technical capability and instead account for how students interpret acceptable use and experience its consequences. The emotional patterns observed in our study indicate that awareness of learning-related and academic risks does not necessarily correspond to heightened emotional engagement. This suggests that future research would benefit from analytical frameworks that integrate affective response with contextual pressure and interpretive uncertainty, particularly within cumulative and practice-oriented disciplines such as software engineering.

Our findings also point to the need for activity-sensitive perspectives in both research and educational analysis. Students in our study associated inappropriate LLM use with a range of software engineering activities, including programming, debugging, testing, documentation, system design, and group work. Treating LLM-assisted cheating as a uniform behavior risks hiding meaningful variation across these activities. Educational designs that differentiate between stages of the software engineering process may provide more accurate accounts of where boundary crossing occurs and how it relates to learning outcomes.

Finally, our results indicate that educational responses centered solely on prohibition or enforcement may be misaligned with students’ lived experiences. Participants described LLM-assisted cheating as embedded in routine coursework and shaped by sustained workload, deadlines, and ambiguous guidance, rather than by lack of awareness of consequences. Our implications for software engineering education, therefore, include the importance of clearer alignment between assessment design, instructional guidance, and learning objectives, as well as preserving opportunities for sustained engagement with core engineering practices that students themselves associate with skill development and professional preparedness.

\subsection{Threats to Validity}

\textbf{Construct validity.} This study relied on self-reported accounts from participants who identified their own LLM use as inappropriate, introducing risks related to recall bias, social desirability, and variation in how academic integrity expectations were interpreted~\cite{pfleeger2001principles,linaker2015guidelines,ralph2020empirical}. Participants may also have differed in how they interpreted concepts such as inappropriate use or misuse depending on institutional policies and instructor expectations. Some survey questions intentionally allowed interpretive flexibility to capture reflective experiences across different coursework and assessment contexts, which may have resulted in variation in how participants interpreted specific prompts. These risks were mitigated through anonymous data collection, piloting, screening questions, and multiple data quality checks, including attention checks and manual review of open-ended responses. \textbf{Internal validity.} The qualitative analysis involved interpretive coding and categorization of participants' narratives, which may introduce researcher subjectivity. To reduce this risk, coding was conducted by at least two researchers, with discrepancies discussed and resolved through consensus. The affective interpretation using the SAM framework also involved interpretive judgment when distinguishing valence and arousal dimensions from written narratives. To improve consistency, emotional interpretations were grounded in participants' explicit language and contextual descriptions. \textbf{External validity.} The study used non-probability sampling and a sample size intended to support analytical rather than statistical generalization. Although participants represented multiple institutional contexts and different stages of software engineering programs, the sample may not reflect the experiences of all software engineering students or institutions. Consequently, the findings should be interpreted as descriptive of students' reported experiences within contemporary software engineering education rather than as representative of all software engineering students or institutional contexts.
\section{Conclusions and Future Work}
\label{sec:conclusion}

This paper presented an empirical account of how software engineering students describe their emotional experiences after using large language models in ways they perceive as academically inappropriate. Based on a cross-sectional survey of 116 undergraduate students, the findings suggest a pattern of emotional attenuation surrounding inappropriate LLM use. Across contexts, emotions were frequently described as neutral, mixed, or situational, even when participants acknowledged learning related or academic risks. Indifference was not associated with ignorance of consequences, but instead appeared alongside awareness of potential harm, suggesting that emotional disengagement may help accommodate routine boundary crossing within everyday coursework. When negative emotions such as guilt or anxiety were reported, they typically reflected internal moral discomfort or concern about institutional consequences rather than sustained emotional strain. Positive emotions, including relief and occasional satisfaction, further indicate that emotionally reassuring responses can coexist with acknowledged rule violations. These findings suggest that emotionally muted or ambivalent responses are an important aspect of how software engineering students experience academically inappropriate LLM use, challenging assumptions that integrity violations are necessarily accompanied by strong negative affect.

In future work, we will investigate how emotional experiences vary across software engineering topics, levels of seniority, activity types, and educational contexts. We also plan to complement survey data with interviews involving both students and instructors, providing richer accounts of how emotional responses are shaped by curricular structures, assessment practices, and instructional guidance. This will support the development of empirically grounded strategies for integrating LLMs into software engineering education while addressing academic integrity concerns in ways that remain attentive to learning goals, disciplinary practices, and students’ lived experiences.

\section*{Data Availability}
\label{sec:DataAvailability}

The data analyzed in this research is available at: \url{https://figshare.com/s/29c372b1260517aa0eb3}.

\begin{credits}

\subsubsection{\discintname}
The authors have no competing interests to declare that are relevant to the content of this article.

\end{credits}

\bibliographystyle{splncs04}
\bibliography{bib}

\end{document}